\documentstyle[preprint,aps]{revtex}
\begin{document}
\draft
\preprint{}
\title{Quantum field theory of cooperative atom response:\\Low
light intensity}
\author{Janne Ruostekoski and Juha Javanainen}
\address{
Department of Physics, University of Connecticut, Storrs,
Connecticut 06269-3046}
\date{\today}
\maketitle
\begin{abstract}
We study the interactions of a possibly dense and/or quantum
degenerate gas with driving light. Both the atoms and the
electromagnetic fields are represented by quantum fields
throughout the analysis. We introduce a field theory version of
Markov and Born approximations for the interactions of light
with matter, and devise a procedure whereby certain types of
products of atom and light fields may be put to a desired,
essentially normal, order. In the limit of low light intensity we
find a hierarchy of equations of motion for correlation functions
that contain one excited-atom field and one, two, three, etc.,
ground state atom fields. It is conjectured that the entire
linear hierarchy may be solved by solving numerically the classical
equations for the coupled system of electromagnetic fields and
charged harmonic oscillators. We discuss the emergence of
resonant dipole-dipole interactions and collective linewidths, and
delineate the limits of validity of the column density approach in
terms of non-cooperative atoms by presenting a mathematical example
in which this approach is exact.
\end{abstract}

\narrowtext
\section{Introduction}\label{sec:intro}
Indications of Bose-Einstein condensation (BEC) in trapped
alkali metal vapors have been reported recently
\cite{AND95,BRA95,DAV95}. At this point all direct probing of such
condensates has been carried out optically. Correspondingly, in
anticipation of BEC and the role of light in the experiments, optical
response of degenerate atomic gases has been the subject of active
theoretical research already for quite some time
\cite{SVI90,POL91,LEW93,JAV94,LEW94,MOR95,JAV95,LEW96,GRA96}. Aside
from the interest in BEC, in an evaporatively cooled gas of alkali
atoms one may for the first time have a homogeneously broadened,
weakly interacting system at such a high density that there are many
atoms in a cubic wavelength, $\rho\lambda^3 \gg 1$. This kind of a
sample would in its own right serve to further our understanding of
the interactions of light with matter.

Nonetheless, in spite of all the theoretical work, there still are
quite basic unsettled issues in the theory of the optical properties of
dense and/or degenerate gases. Under the condition
$\rho\lambda^3 \gg 1$ the atoms no longer respond to the
electromagnetic fields individually, but their properties are modified
by the presence of nearby atoms. For instance, the atoms exhibit
collective linewidths and line shifts. Inasmuch as it comes to the
near-resonant response in the regime $\rho\lambda^3 \gg 1$, all
treatments of the optical properties of degenerate gases known to the
present authors (including ours) contain uncontrolled approximations
that bear on linewidths and line shifts. As a result, the regions of
validity and the relations between different treatments tend to be
somewhat ill-defined.

A rigorous study of atom-field interactions valid regardless
of atom density, atom statistics, optical detuning, and so forth, is
clearly called for. The paper of Morice {\it et al}. \cite{MOR95} is a
step in this direction. They start from a full quantum mechanical
Hamiltonian, including quantized light and internal degrees of freedom
and c.m.\ motion of the atoms. However, at an early stage these authors
go over to a classical treatment of the c.m.\ motion of the atoms. They
then derive equations of motion for a few correlation functions
involving polarization and atom density, and solve the optical response
including all photon exchange between any pair of atoms.

The  program carried out in Sec.~\ref{sec:gen} of the present paper is
similar to the agenda of Ref.~\cite{MOR95}. We start in
Sec.~\ref{sec:bfm} from our field theory version of the Hamiltonian as
in Ref.~\cite{JAV95}, amended with the atom-atom contact interaction
\cite{COH89,MOR95} that derives from the Power-Zienau approach. The
point of departure from Ref.~\cite{MOR95} is that we describe the
atoms with quantum fields throughout. The mathematical techniques
introduced in the process are analogous to the time honored tools
in quantum optics: field theory version of the Born-Markov
approximation (Sec.~\ref{sec:vac}), and procedures to move
noncommuting operators to a certain advantageous order
(Sec.~\ref{sec:hie}). In the present paper we complete the
derivation by assuming the limit of low intensity for driving
light. The end result in Sec.~\ref{sec:cor} is a hierarchy of
equations of motion for correlation functions that involve atomic
polarization at one point and densities at 0, 1, \ldots, points in
space. The lowest two equations coincide with those given in
Ref.~\cite{MOR95}.

In the present paper the emphasis is on the structure of the theory.
To gain more insights, we examine a few simple special cases in
Sec.~\ref{sec:exa}. We demonstrate the exquisite subtlety of the
propagation of radiation through an atomic sample by presenting one
particular set of assumptions that yields the standard column density
results of (optically) non-cooperating atoms (Sec.~\ref{sec:sib}), and
by reviewing the density expansion of Morice {\it et al}.
(Sec.~\ref{sec:det}). In the case of only two atoms, what we call
cooperative linewidth and line shift emerge as manifestation of
the dipole-dipole interaction. This is discussed in
Sec.~\ref{sec:rdd}. Here we also point out that in our limit of low
intensity of the driving light, the collective linewidth and line
shift could perfectly well have been derived from classical
electrodynamics of classical atoms (charged harmonic oscillators).
In this paper we do not attempt to derive any new results from the
hierarchy of correlation functions, but the connection to classical
physics points to a possible future method for exact solution of the
hierarchy: classical simulations of a system of classical atoms. A
few comments to this effect are made in Sec.~\ref{sec:pfe}.

Concluding remark about the possible solutions and extensions of
our hierarchy are made in the final Sec.~\ref{sec:con}. Certain
mathematical details concerning the divergence of the dipolar
field and a summary of dipole matrix elements are deferred
to the Appendix.

\section{Motion of atom fields}\label{sec:gen}
\subsection{Basic dynamics of the fields}\label{sec:bfm}

We begin by recapping, reformulating, and extending the salient
results of \cite{JAV95}. The main items of this section are the coupled
evolution equations for the light and matter fields exemplified by
Eqs.~(\ref{eq:FEQ}) and~(\ref{eq:MEQ}). Overall, we emphasize the
similarities of the theory to the classical electrodynamics of
polarizable media.

\subsubsection{Hamiltonian}\label{seq:HAM}
For better or worse, in this paper we regard atoms as point dipoles.
A mathematically rigorous treatment produces a delta function
term in the field of a dipole at the position of the dipole,
which results in a contact interaction between dipoles. For
mathematical consistency,  this time around we therefore also
keep the contact interaction generated in the Power-Zienau
transformation from the
${\bf p}\cdot{\bf A}$ to the ${\bf d}\cdot{\bf E}$ gauge 
\cite{COH89,MOR95}.
This interaction was ignored as presumably inconsequential in
the limit of large detuning considered in Ref.~\cite{JAV95}, but for an
arbitrary detuning it may become an issue.

The atoms have  two internal energy levels, which we label
$g$ for ``ground'' and $e$ for ``excited''. We allow for
angular momentum degeneracy of the energy levels, so the complete
specification of the internal state of an atom $\alpha m$ includes the
level label $\alpha$ = $e$ or $g$  and the $z$ component of angular
momentum $m$. We assume dipole coupling of each atom to light.

In first quantization, we add to the Hamiltonian of \cite{JAV95} the
contact interaction, the polarization energy
\begin{equation} H_{\rm P} = {1\over2\epsilon_0}
\sum_{i\ne j} {\bf d}_i\cdot{\bf d}_j\,\delta({\bf r}_i-{\bf r}_j)\,.
\label{eq:PE1}
\end{equation}
Here ${\bf d}_i$ and ${\bf r}_i$ are the dipole operator and
center-of-mass position operator for the $i^{\rm th}$ atom. There are
also divergent self-energies with $i=j$, but we ignore these as we do
not attempt a quantitative calculation of the Lamb shift. Equation
(\ref{eq:PE1}) displays a standard two-body interaction, which is
immediately converted to second quantization. As before, the many-atom
system is described by Heisenberg picture quantum fields
$\psi_{\alpha m}({\bf r} t)$, which obey the proper commutator
relations. While much of our development applies to fermions as well,
in this paper we consider only bosons explicitly. In terms of the atom
fields, the additional polarization energy is the integral of the
Hamiltonian density
\begin{eqnarray} {\cal H}_{\rm P} &=& {1\over\epsilon_0}
\sum_ {\renewcommand{\arraystretch}{0.2}
\begin{array}{cc}
\scriptstyle m_1 m_2\\ \scriptstyle M_1M_2
\end{array}
\renewcommand{\arraystretch}{1.0}} [{\bf d}_{m_2M_1}\cdot{\bf d}_{M_2m_1}\,
\psi^\dagger_{gm_2}\psi^\dagger_{eM_2}
\psi_{eM_1} \psi_{gm_1}
\nonumber\\ &+& \frac{1}{2}\,{\bf d}_{M_2m_2}\cdot{\bf d}_{M_1m_1}\,
\psi^\dagger_{eM_2}\psi^\dagger_{eM_1}
\psi_{gm_2} \psi_{gm_1} +\frac{1}{2}\,{\bf d}_{m_2M_2}\cdot
{\bf d}_{m_1M_1}\,
\psi^\dagger_{gm_2}\psi^\dagger_{gm_1}
\psi_{eM_2} \psi_{eM_1}]\,.
\label{eq:PE21}
\end{eqnarray}
The notation ${\bf d}_{mM}$ stands for the matrix element
$\langle gm | {\bf d} | eM\rangle$ of the dipole operator of one atom,
${\bf d}$. We denote the energy level implicitly in such a way that a label
of a Zeeman state with a lower-case $m$ refers to ground state,
upper-case
$M$ to the excited state.

However, to simplify the notation as far as possible, we are going to
adopt yet another convention that is in force unless we explicitly
state otherwise. We do not write the magnetic quantum numbers
explicitly. For instance, we write
$\psi_{e_2}$ in lieu of
$\psi_{eM_2}$. Also, we write the matrix elements ${\bf d}_{mM}$ as
${\bf d}_{ge}$. Finally, if the same level index appears twice in a
product, sum over the magnetic substates of the level is implied. With
these conventions, we write Eq.~(\ref{eq:PE21}) anew as
\begin{eqnarray}
{\cal H}_{\rm P} &=& {1\over\epsilon_0}
[{\bf d}_{g_2e_1}\cdot{\bf d}_{e_2g_1}
\psi^\dagger_{g_2}\psi^\dagger_{e_2}
\psi_{e_1} \psi_{g_1}
+\frac{1}{2}\,{\bf d}_{e_2g_2}\cdot{\bf d}_{e_1g_1}\,
\psi^\dagger_{e_2}\psi^\dagger_{e_1}
\psi_{g_2} \psi_{g_1}\nonumber\\
&+& \frac{1}{2}\,{\bf d}_{g_2e_2}\cdot
{\bf d}_{g_1e_1}\,
\psi^\dagger_{g_2}\psi^\dagger_{g_1}
\psi_{e_2} \psi_{e_1}]\,.
\end{eqnarray}

\subsubsection{Electromagnetic fields}\label{sec:EMF}
Unlike in \cite{JAV95}, and similarly to \cite{MOR95}, we assume that
there is a cutoff in the wave numbers $q$ of the photons; we multiply
the density of the states of the electromagnetic fields by
$e^{-q^2\alpha^2/4}$, with
$\alpha > 0$ being a length scale. The cutoff removes all mathematical
problems concerning, e.g., the exchange of the order of derivatives
and integrals, which are abundant in the theory without the cutoff. At
the end of the calculations we ultimately take the limit
$\alpha\rightarrow0$.

In spite of the change in the Hamiltonian and the added cutoff of
photon frequencies, the analysis of the electromagnetic fields proceeds
almost as in \cite{JAV95}. In accordance with Ref.~\cite{COH89}, it
emerges from our results that what was called the electric field in
\cite{JAV95} should more properly be interpreted as the electric
displacement divided by the permittivity of the vacuum $\epsilon_0$. We
henceforth adopt this interpretation. The positive frequency part of
the electric displacement is expressed in terms of the matter fields as
\begin{equation}
{\bf D}^+({\bf r} t) = {\bf D}^+_F({\bf r} t) +
\epsilon_0\int_{-\infty}^tdt'\,
\int d^3r'\,{\bf S}({\bf d}_{ge};{\bf r}-{\bf r'},t-t')\,
\psi^\dagger_{g}({\bf r'} t'){\psi}^{{\rule[0.3ex]{0pt}{0pt}}}_{e}({\bf r'} 
t')\,.
\label{eq:FEQ}
\end{equation}
In this approach electric displacement and matter fields are the
primary degrees of freedom; ${\bf D}^+_F({\bf r} t)$ is the free
electric displacement that would apply if there were no coupling
between matter and electromagnetic fields. The
propagator that takes the radiation from the dipole source to the
field point is
\begin{mathletters}
\begin{eqnarray}
{\bf S}(\mbox{\boldmath$\cal D$};{\bf r} t) &=& 
{ic\over16\pi^3\epsilon_0}
\int d^3q\, e^{-q^2\alpha^2/4}\,
q\,{{\bf q}\over q}\times\left( {{\bf q}\over q}\times
\mbox{\boldmath$\cal D$}\right)
e^{i{\bf q}\cdot {\bf r}}(e^{icqt}-e^{-icqt})\,
\label{eq:SR0} \\
&=& 
{c\over 4\pi\epsilon_0}
\,(\mbox{\boldmath$\cal D$}\!\times\!\mbox{\boldmath$\nabla$})\!\times\!
\mbox{\boldmath$\nabla$}\,
{\delta_\alpha[r-ct] - \delta_\alpha[r+ct]
\over |{\bf r}-{\bf r'}|}\,,
\label{eq:SR1}
\end{eqnarray}
\end{mathletters}
where the delta function has acquired  a finite width as a result of
the cutoff in the photon energy spectrum:
\begin{equation}
\delta_\alpha(x) = {1\over\sqrt\pi\alpha}\,
\exp\left[{-\left({x\over\alpha}\right)^2}\right]\,.
\label{eq:DWT}
\end{equation}

As before, we assume that there is a dominant frequency
$\Omega$ in the problem. In fact, we generally assume
that a field such as $\psi_{e}({\bf r} t) e^{i\Omega t}$, and similarly
for the electromagnetic quantities, varies ``slowly'' in time in
comparison with $e^{-i\Omega t}$. From now on a notation such as
$\psi_{e}({\bf r} t)$ refers to the slowly varying field
$\psi_{e}({\bf r} t) e^{i\Omega t}$, unless explicitly stated otherwise.

Based on an implicit cutoff such as $\alpha$, we argued in \cite{JAV95}
that the delta functions with plus and minus signs in Eq.~(\ref{eq:SR1})
conspire to remove a term $\propto\delta({\bf r}-{\bf r'})$ that
results when the position derivatives act on $|{\bf r}-{\bf r'}|^{-1}$.
What we did not realize is that this delta function does not outright
vanish. Instead, it gets smeared to a function whose integral over
${\bf r}$ is still unity but which has a finite width of the order
$\alpha$; see Appendix
\ref{app:CMT}. From now on, as long as the integral operator with
the kernel $\bf S$ acts on any smooth function $\phi({\bf r}
t)e^{-i\Omega t}$ of ${\bf r}$ and $t$ in which the exponential is the
dominant time dependence, we write
\begin{equation}
\int_{-\infty}^tdt'\, \int d^3r'\,{\bf S}(\mbox{\boldmath$\cal D$};
{\bf r}-{\bf r'},t-t')\phi({\bf r'} t')
e^{-i\Omega t'}
= e^{-i\Omega t}\int d^3r'\,{\bf S}'(\mbox{\boldmath$\cal D$};
{\bf r}-{\bf r'}) \phi({\bf r'} t_d)\,.
\label{eq:SMI}
\end{equation}
Here 
\begin{equation}
t_d = t-{|{\bf r}-{\bf r'}|\over c}
\label{eq:RET}
\end{equation}
is the retarded time. The monochromatic version of the propagator ${\bf
S}$, ${\bf S}'$, may be written alternatively
\begin{mathletters}
\begin{equation}
{\bf S}'(\mbox{\boldmath$\cal D$};{\bf r}) = {1\over4\pi\epsilon_0}
(\mbox{\boldmath$\cal
D$}\times\mbox{\boldmath$\nabla$})\times\mbox{\boldmath$\nabla$}\,
{e^{ikr}\over r}\,,
\label{eq:KR1}
\end{equation}
or
\begin{equation}
{\bf S}'(\mbox{\boldmath$\cal D$};{\bf r}) = {\bf
K}(\mbox{\boldmath$\cal D$};{\bf r}) +
{2\over3\epsilon_0}\,\mbox{\boldmath$\cal D$}\,\delta({\bf r})\,.
\label{eq:KR2}
\end{equation}
\label{eq:KRA}
\end{mathletters}
The final new kernel $\bf K(\mbox{\boldmath$\cal D$};{\bf r})$ is equal
to the positive-frequency component of the electric field from a
monochromatic dipole with the complex amplitude $\mbox{\boldmath$\cal
D$}$, given that the dipole resides at origin and the field is observed
at
$r\ne0$. The explicit expression is, of course
\cite{JAC75},
\begin{equation} {\bf K}(\mbox{\boldmath$\cal D$};{\bf r}) =
{1\over4\pi\epsilon_0}
\bigl\{ k^2(\hat{\bf n}\!\times\!\mbox{\boldmath$\cal
D$})\!\times\!\hat{\bf n}{e^{ikr}\over r} +[3\hat{\bf n}(\hat{\bf
n}\cdot\mbox{\boldmath$\cal D$})-\mbox{\boldmath$\cal D$}]
\bigl( {1\over r^3} - {ik\over r^2}\bigr) e^{ikr}
\bigr\},
\label{eq:DOL}
\end{equation}
with
\begin{equation}
\hat{\bf n} = {{\bf r}\over r}, \qquad k = {\Omega\over c}\,.
\label{eq:NOT}
\end{equation}
It should be noted that, as the dipole radiation has the $1/r^3$
singularity, integrals such as (\ref{eq:SMI}) are generally not
absolutely convergent. According to Appendix
\ref{app:CMT}, we resolve this problem by the rule that at least in
the immediate vicinity of the divergence ${\bf r'}={\bf r}$ the
integral is to be done in spherical polar coordinates, and the angles
are to be integrated over first. It should also be born in mind that the
form (\ref{eq:SMI}) does not apply if the function $\phi$ is singular in
{\bf r}.

We finally consider the quantum expectation value of
Eq.~(\ref{eq:FEQ}). As is always done in this paper, we take the free
field to be in a coherent state. We also assume that the expectation
value
$\langle{\bf D}^+_F\rangle$ is effectively monochromatic. It is physically
evident that, at least in steady state, the expectation value of the
product $\psi^\dagger_g\psi_e$ will then be monochromatic and a
smooth function of ${\bf r}$ as well. Inasmuch as the quantum
expectation value of Eq.~(\ref{eq:FEQ}) is concerned, the transformation
from the kernel
$\bf S$ to the kernel
$\bf S'$ is thus allowed. Moreover, the expectation value of the free
field is a solution to the Helmholtz equation, and the function
$e^{ikr}/r$ is essentially the Green's function of the Helmholtz
differential operator:
\begin{equation}
(\mbox{\boldmath$\nabla$}^2+k^2)\langle{\bf D}^+_F\rangle = 0, \quad
(\mbox{\boldmath$\nabla$}^2+k^2) {e^{ikr}\over r} = -4\pi\delta({\bf
r})\,.
\end{equation}
In view of Eqs.~(\ref{eq:SMI}) and~(\ref{eq:KR1}), we thus have from
Eq.~(\ref{eq:FEQ})
\begin{equation}
(\mbox{\boldmath$\nabla$}^2+k^2)\langle{\bf D}^+\rangle =
-\mbox{\boldmath$\nabla$}\times(\mbox{\boldmath$\nabla$}
\times\langle{\bf P}^+\rangle)\,.
\label{eq:FEX}
\end{equation}
Classically, the polarization of the medium is defined as dipole
moment per atom times the density of atoms. It is therefore clear that
\begin{equation}
{\bf P}^+({\bf r}) =
{\bf d}_{ge}\psi^\dagger_{g}({\bf r})\psi_{e}({\bf r})
\label{eq:PDF}
\end{equation}
should be identified as the (positive frequency part of the) quantum
mechanical polarization operator.

The value of Eq.~(\ref{eq:FEX}) is twofold. First, it is a local
differential equation, as opposed to the integral equation
(\ref{eq:FEQ}). Second, it has a well-known counterpart in the classical
electrodynamics of polarizable media. This reinforces the
interpretations of ${\bf D}$ and $\bf P$ as electric displacement and
polarization operators.

\subsubsection{Matter fields}\label{sec:MAF}
We now turn to the equations of motion of the matter fields.
Under the assumptions that the density of excited atoms is low and
that an atom moves much less than a wavelength of resonant light
during the time it remains excited, we have the equations of motion
for the fields describing excited and ground state atoms,
\begin{mathletters}
\label{eq:MEQ}
\begin{eqnarray}
\dot{\!\psi}_{e}({\bf r} t) &=&  i \delta\,{\psi}_{e}({\bf r} t) +
{i\over\hbar}\,{\bf d}_{eg}\cdot{\bf E}^+({\bf r} t)\,\psi_{g}({\bf r}
t)\,,
\label{eq:EXF}\\
\dot{\psi}_{g}({\bf r} t) &=&  {i\over\hbar}\,
{\bf E}^-({\bf r} t)\cdot{\bf d}_{ge}\,{\psi}_{e}({\bf r} t)
-i\,{H_{\hbox{\scriptsize c.m.}}^{{\rule[0.3ex]{0pt}{0pt}}}({\bf
r})\over\hbar}\,{\psi}_{g}({\bf r} t) +\left.{d\over
dt}\right|_{\hbox{\scriptsize C}}\psi_{g}({\bf r} t)\,.
\label{eq:GRF}
\end{eqnarray}
\label{eq:MAT}
\end{mathletters}
As the notation implies,
\begin{equation}
{\bf E}^+({\bf r} t) = {1\over\epsilon_0}\,[{\bf D}^+({\bf r} t) -  {\bf
P}^+({\bf r} t)]
\label{eq:DEL}
\end{equation}
is to be interpreted as the electric field. Furthermore, $\delta =
\Omega - \omega_0$ is the detuning of the characteristic frequency of
the light $\Omega$ from the atomic resonance frequency
$\omega_0$. We have carried out the Rotating Wave Approximation that
takes into account the dominant field frequency $\Omega$. Finally,
$H_{\hbox{\scriptsize c.m.}}^{{\rule[0.3ex]{0pt}{0pt}}}$ is the one-particle Hamiltonian governing the c.m.\  motion
of the atoms, and the time derivative with the subscript C
represents collisions.

\subsubsection{Summary remark}\label{sec:SRM}

The two-state model of quantum optics is immediately seen to
underlie Eqs.~(\ref{eq:MAT}), and Eq.~(\ref{eq:FEQ}) describes the
total field as the incident field plus the fields radiated by the
dipoles, complete with propagation delays. In spite of the familiar
appearances, though, it should be noted that the only real
approximations so far have been to ignore the c.m.\ motion and
collisions of the excited atoms. The formulation still fully accounts
for quantum statistics of the many-atom system, and for the quantized
electromagnetic fields. The effect of the dipole-dipole interactions on
the transition frequencies and linewidths of the atoms is still
included. On the other hand, as we have ignored the c.m.\ Hamiltonian
for excited atoms, collisions between ground state atoms and excited
atoms can no longer be discussed.

\subsection{Eliminating the vacuum field}\label{sec:vac}

Even in the absence of applied electromagnetic fields the atoms bathe
in vacuum fields that cause spontaneous emission and Lamb shifts.
The purpose of the present section is to account for the vacuum
fields. While pursuing this goal, we need to be prepared for singular
functions with rapid spatial and temporal variations. We therefore
start with the general field equation (\ref{eq:FEQ}). Moreover, for the
time being we argue in terms of the original atomic and electromagnetic
fields untempered by the exponential $e^{i\Omega t}$.

To begin with, we insert Eq.~(\ref{eq:FEQ}) into Eq.~(\ref{eq:EXF}), and
obtain
\begin{eqnarray}
\lefteqn{\dot{\!\psi}_{e}({\bf r} t) =   -i\omega_0\,{\psi}_{e}({\bf r}
t)}
\nonumber\\ &+& {i\over\hbar\epsilon_0}
{\bf d}_{eg}\cdot 
\left\{\rule{0ex}{3ex}\right.
{\bf D}^+_F({\bf r} t)\psi_{g}({\bf r} t)
-{\bf d}_{g'e'}
\psi^\dagger_{g'}({\bf r} t)\psi_{g}({\bf r} t)\psi_{e'}({\bf r} t)
\nonumber\\
&+& 
\epsilon_0\int_{\infty}^tdt'\int d^3r'\,
{\bf S}({\bf d}_{g'e'};{\bf r}-{\bf r'}, t-t')
\psi^\dagger_{g'}({\bf r'} t')
{\psi}^{{\rule[0.3ex]{0pt}{0pt}}}_{e'}({\bf r'} t')
\psi_{g}({\bf r} t)
\left.\rule{0ex}{3ex}\right\}\,.
\label{eq:SP1}
\end{eqnarray}

We are ultimately interested in quantum expectation values of
atomic and electromagnetic field operators, and thus wish to be able to
take expectation values of expressions such as Eq.~(\ref{eq:SP1})
easily. It would be especially valuable to have the free-field operator
${\bf D}^+_F({\bf r} t)$ farthest to the right. Because the (initial)
free electromagnetic field is assumed to be in the coherent state, in an
expectation value this operator would then reduce to a
multiplicative classical field: the relation
\begin{equation}
\langle{\cal O}{\bf D}^+_F({\bf r} t)\rangle =
\langle{\cal O}\rangle \langle{\bf D}^+_F({\bf r} t)\rangle
\label{eq:FCT}
\end{equation}
applies to any operator $\cal O$. Evidently we need commutators
between atom operators and free-field operators, so that we
may move the latter to the desired positions.

Any atom operator, of course, commutes with the total electric
displacement operator ${\bf D}^\pm$ at the same time. We thus have
from Eq.~(\ref{eq:FEQ})
\begin{eqnarray}
Q &=& [\psi_{g}(\tilde{\bf r} t),{\bf D}^+_F({\bf r} t)]\nonumber\\
&=& - \epsilon_0\int_{-\infty}^tdt'\,
\int d^3r'\,{\bf S}({\bf d}_{g'e};{\bf r}-{\bf r'},t-t')
[\psi_{g}(\tilde{\bf r} t),
\psi^\dagger_{g'}({\bf r'} t'){\psi}^{{\rule[0.3ex]{0pt}{0pt}}}_{e}
({\bf r'} t')]\,.
\label{eq:CM1}
\end{eqnarray}
Here we are preparing for the eventuality that the commutator is
required for two different field points.

The standard way of dealing with vacuum fields in quantum optics is
the duo of Born and Markov approximations: the atom operators evolve
as if no electromagnetic fields were present (Born) during the short
vacuum correlation time (Markov) \cite{COH77}. For implementations of
this idea in the Heisenberg picture see, e.g.,
\cite{LEH70,ACK74,KIM76}.  We evaluate the commutator $Q$ under an
approximation which, we think, is the field theory equivalent of the
standard Born and Markov approximations: We assume that during the time
it takes radiation reaction effects to assert themselves, the
atom fields evolve as if they were completely noninteracting. We
temporarily restore the explicit notation for magnetic quantum
numbers, and write the Born-Markov approximation for
Eq.~(\ref{eq:CM1}) as
\begin{equation}
\psi_{gm}({\bf r} t') =
{1\over\sqrt V} \sum_{{\bf k}} e^{i[{\bf k}\cdot{\bf r}-\epsilon_{{\bf
k}}(t'-t)]} b_{gm{\bf k}}(t),
\quad
\psi_{eM}({\bf r} t') =
{1\over\sqrt V} \sum_{{\bf k}} e^{i[{\bf k}\cdot{\bf r}-\omega_0
(t'-t)]} b_{eM{\bf k}}(t)\,.
\label{eq:FCM}
\end{equation}
Here $\epsilon_{{\bf k}} = \hbar{\bf k}^2/2m$ gives the dispersion
relation for an atom with mass $m$, $b$ are boson operators, and the
sums run over the wave vectors
${\bf k}$ appropriate for the quantization volume $V$. In the standard
continuum limit the relevant commutator becomes
\begin{equation}
[\psi_{gm}(\tilde{\bf r} t),\psi^\dagger_{gm''}({\bf r'} t')] =
{\delta_{mm''}\over(2\pi)^3}
\int d^3k\,e^{i{\bf k}\cdot(\tilde{\bf r}-{\bf
r'})-i\epsilon_k(t-t')}\,.
\label{eq:EQC}
\end{equation}
We use $\bf S$ from (\ref{eq:SR0}), the commutator from (\ref{eq:EQC}),
$\psi^\dagger_{gm'}$ from (\ref{eq:FCM}), and add the conventional
convergence factor $e^{-\eta t}$ to the time integral.
Equation (\ref{eq:CM1}) is cast into the form
\begin{eqnarray}
\lefteqn{Q = {-ic\over16\pi^3\sqrt{V}}\sum_{{\bf K},M'}
e^{i{\bf K}\cdot\tilde{\bf r}}\,b_{eM'{\bf K}}\,
\int_0^\infty d\tau\,e^{-\eta\tau}\int d^3q}\nonumber\\
&&e^{i{\bf q}\cdot(\tilde{\bf r}-{\bf r})}\,e^{-\alpha^2q^2/4}\,
q\,{{\bf q}\over q}\times\left( {{\bf q}\over q}\times{\bf d}_{mM'}
\right)
(e^{icq\tau}-e^{-icq\tau})\,e^{-i\epsilon_{{\bf K}-{\bf q}}
\tau+i\omega_0\tau}\,.
\end{eqnarray}

Our final approximation is to ignore the c.m.\ energies
in comparison with the energy of the internal transition of the atom;
we write $\omega_0-\epsilon_{{\bf K}-{\bf q}}\simeq\omega_0\simeq\Omega$.
The results is interesting:
\begin{equation}
Q = -\epsilon_0\int_0^\infty d\tau\,
{\bf S}({\bf d}_{ge};\tilde{\bf r}-{\bf r},\tau)e^{i\Omega\tau}\,
\psi_{e}(\tilde{\bf r} t)\,.
\end{equation}
The time integral is the same as the definition of the
kernel ${\bf S}'({\bf d}_{ge};\tilde{\bf r}-{\bf r})$ in
(\ref{eq:KRA}), albeit still containing the cutoff parameter $\alpha$.
The cutoff is truly needed: in our immediate application to
Eq.~(\ref{eq:SP1}) we are to set $\tilde{\bf r}={\bf r}$, and without
the cutoff we would have to contend with a pernicious singularity of the
type $\delta({\bf r})/r^3$. For a small but nonzero
$\alpha$, the result is
\begin{equation}
[\psi_{g}({\bf r} t),{\bf D}^+_F({\bf r} t)] = 
{\bf d}_{ge} \left(
{2\pi^{-3/2}\over3\alpha^3}-i{\omega_0^3\over6\pi c^3} \right)
\psi_{e}({\bf r} t)\,.
\end{equation}

Given the sum rule for the dipole moment matrix elements,
(\ref{eqa:DSM}), the relevant terms in Eq.~(\ref{eq:SP1}) become
\begin{eqnarray}
\dot{\!\psi}_{e}({\bf r} t) &=&   -i\omega_0\,{\psi}_{e}({\bf r} t)
+{i\over\hbar\epsilon_0}\,{\bf d}_{eg}\cdot{\bf D}^+_F({\bf r} t)\,
\psi_{g}({\bf r} t)\ldots
\nonumber\\
&=& -i\omega_0\,{\psi}_{e}({\bf r} t)
 -\left[\gamma +  i{2{\cal D}^2\sqrt\pi\over3\pi^2\epsilon_0\hbar\alpha^3}
\right] \,\psi_{e}
+{i\over\hbar\epsilon_0}\,
{\bf d}_{eg}\cdot\psi_{g}({\bf r} t)\,{\bf D}^+_F({\bf r} t)\ldots\,,
\label{eq:SPT}
\end{eqnarray}
where ${\cal D}$ is the reduced dipole moment matrix element. The imaginary
part in the second term on the right-hand side of
Eq.~(\ref{eq:SPT}) diverges as the photon momentum cutoff goes to
infinity with
$\alpha\rightarrow0$. This part, after a proper renormalization,
contributes to the Lamb shift. From now on we assume that the Lamb
shift is already included in the transition frequency, and ignore the
$\alpha^{-3}$ term in (\ref{eq:SPT}). What remains is the familiar
spontaneous linewidth of the atomic transition,
\begin{equation}
\gamma = {{\cal D}^2\omega_0^3\over6\pi\hbar\epsilon_0 c^3}\,.
\end{equation}

There are no divergence problems with the commutator of $\psi_{g}$ and
${\bf D}^+_F$ if the position arguments are different. We simply write
\begin{mathletters}
\begin{equation}
[\psi_{g}(\tilde{\bf r} t),{\bf D}^+_F({\bf r} t)] =
-\epsilon_0\,{\bf S}'({\bf d}_{ge};{\bf r}-\tilde{\bf
r})\psi_{e}(\tilde{\bf r} t)\,.
\label{eq:RRF}
\end{equation}
In fact, if the divergent in-phase part of the dipole field at
$\tilde{\bf r}={\bf r}$ is ignored (or incorporated in the Lamb shift),
we may interpret Eq.~(\ref{eq:RRF}) to be valid even for $\tilde{\bf
r}={\bf r}$.

By the same token, we may carry out all commutators between atom
fields and free electromagnetic fields. The two additional
nonvanishing commutators that play some role in this paper are
\begin{eqnarray}
[ \psi^\dagger_{g}(\tilde{\bf r} t),{\bf D}^-_F({\bf r} t)] &=&
\epsilon_0\psi^\dagger_{e}(\tilde{\bf r} t)
[{\bf S}'({\bf d}_{ge};{\bf r}-\tilde{\bf r})]^*\,,
\label{eq:URF}\\
\hbox{[}\psi_{e}(\tilde{\bf r} t),{\bf D}^-_F({\bf r} t)]&=&
-\epsilon_0\psi_{g}(\tilde{\bf r} t)
[{\bf S}'({\bf d}_{ge};{\bf r}-\tilde{\bf r})]^*\,.
\label{eq:VRF}
\end{eqnarray}
\label{eq:CMS}
\end{mathletters}

At this point we restore our convention of slowly varying fields. We
also add another assumption to the effect that light has ample time to
propagate across the atomic sample during the time that the slowly
varying fields change appreciably. This permits us to ignore
propagation delays in the time arguments of the slowly varying fields.
The retarded time $t_d$ is simply replaced by the external time $t$.
We have thus obtained an equation of motion for the excited state
atom field that contains an explicit radiative damping,
\begin{eqnarray}
\dot{\!\psi}_{e}({\bf r}) &=&   (i \delta-\gamma)
{\psi}_{e}({\bf r})
\nonumber\\
&+& {i\over\hbar\epsilon_0}\,
{\bf d}_{eg}\cdot 
\left\{\rule{0ex}{3ex}\right.
\psi_{g}({\bf r}){\bf D}^+_F({\bf r}) -{\bf d}_{g'e'}
\psi^\dagger_{g'}({\bf r})\psi_{e'}({\bf r})\psi_{g}({\bf r})
\nonumber\\ &+& 
\epsilon_0\int d^3r'\, {\bf S}'({\bf d}_{g'e'};{\bf r}-{\bf r'})
\psi^\dagger_{g'}({\bf r'})
{\psi}^{{\rule[0.3ex]{0pt}{0pt}}}_{e'}({\bf r'})\psi_{g}({\bf r})
\left.\rule{0ex}{3ex}\right\}\,.
\label{eq:SPE}
\end{eqnarray}
Here, and in our subsequent expressions, the common time $t$ is omitted
in the notation.

Unlike in the ordinary treatments of spontaneous emission in
the quantum optics of an isolated atom, no short vacuum correlation
time suggests itself in our formulation. The use of the free-field
evolution as in (\ref{eq:FCM}) during the ``vacuum correlation time''
may thus seem like an ad-hoc assumption. This approximation,
however, did produce spontaneous damping and Lamb shift in
accordance with the one-atom theory. Of course, even in standard
quantum optics the atomic variables do not evolve completely freely
during the vacuum correlation time. Spontaneous emission itself, as
well as external driving electromagnetic fields and collisions
between the atoms in principle affect spontaneous emission, but at
ordinary  conditions for laser spectroscopy these influences are
negligible \cite{COH77}. We conjecture that the same applies in our
field theory version of spontaneous emission. Finally, ignoring
c.m.\  energies in comparison with the energy of the atomic
transition is nothing new either. This is a standard approximation
in the derivation of spontaneous emission in the theory of light
pressure \cite{STE74}. If such energies {\em were\/} included, a
velocity dependent spontaneous emission rate would emerge in
manifest contradiction with special relativity \cite{WIL94}.

\subsection{A hierarchy for operator products}\label{sec:hie}

A particularly relevant atomic variable is the polarization
operator, (\ref{eq:PDF}), which acts as the source for secondary
radiation. Generalizing slightly, we now embark on a study of the time
evolution of the operator product $\psi^\dagger_{g}({\bf
r})\psi_{e}({\bf r})$.

We have in mind situations in which collisions and c.m.\ motion of
the ground state atoms have come to a steady state before the
driving light is turned on. We regard the external field as
a small perturbation, so that the ground state atoms remain materially
unperturbed in the presence of the light. As the final item, we assume
that collisions and c.m.\ motion of the ground state atoms take
place on a time scale much longer than the spontaneous emission time
scale $\gamma^{-1}$. They are therefore not expected to
interfere with spectroscopic probing of the atomic transition. Under
these assumptions we will henceforth ignore the collision terms and the
c.m.\ evolution of the ground state atoms altogether. Nonetheless,
mathematical consistency dictates that in the intermediate steps of our
calculations we take into account some light driven
evolution even for the ground state atoms.

We thus have the equation of motion from Eqs.~(\ref{eq:GRF})
and~(\ref{eq:SPE}),
\begin{eqnarray}
\lefteqn{{d\over dt}\,\psi^\dagger_{g}\psi_{e} =
(i\delta-\gamma)\psi^\dagger_{g}\psi_{e}}\nonumber\\
&& + {i\over\hbar\epsilon_0}\,{\bf d}_{eg'}\cdot
\left\{\rule{0ex}{3ex}\right.
\psi^\dagger_{g}\psi_{g'}{\bf D}^+_F - 
\psi^\dagger_{g}{\bf P}^+\psi_{g'}+
\epsilon_0\int d^3r'\,\psi^\dagger_{g}{\bf S}'({\bf P}^+({\bf r'}),
{\bf r}-{\bf r'})\psi_{g'}
\left.\rule{0ex}{3ex}\right\}\nonumber\\
&& - {i\over\hbar\epsilon_0}\,{\bf d}_{e'g}\cdot
\left\{\rule{0ex}{3ex}\right.
\psi^\dagger_{e'}\psi_{e}{\bf D}^+_F - 
\psi^\dagger_{e'}{\bf P}^+\psi_{e}+
\epsilon_0\int d^3r'\,\psi^\dagger_{e'}{\bf S}'({\bf P}^+({\bf r'}),
{\bf r}-{\bf r'})\psi_{e}
\left.\rule{0ex}{3ex}\right\}\,,
\label{eq:CR1}
\end{eqnarray}
where we have shown explicitly only the nonlocal position
dependence. Generalizing, we are evidently about to derive a hierarchy
of equations for operators of the type ${\bf P}^+$,
$\psi^\dagger_g{\bf P}^+\psi_g$,
$\psi^\dagger_g\psi^\dagger_g{\bf P}^+\psi_g\psi_g$, \ldots, with
different position arguments for the different fields.

Now, light has to be present in order to produce excited
atoms. Each excited state field corresponds to one order in
the strength of the driving light. To first order in ${\bf D}^\pm_F$ we
might thus ignore the second term on the right-hand side of
Eq.~(\ref{eq:CR1}) altogether. This is, generally speaking, what we will
do: we only retain those products of operators that involve at most one
of the operators ${\bf D}^+_F$, ${\bf D}^-_F$, $\psi_e$ or
$\psi^\dagger_e$. However, caution must be exercised for two reasons.
First, we will eventually arrange all atom fields to normal order;
creation operators to the left, annihilation operators to the right.
Besides, we move the free-field operators to prescribed positions as
well. In the process commutators are generated that may be of different
order in the strength of the driving field than the original terms.
Second, some of the commutators are flat out divergent, analogous to
the Lamb shift. It may be shown that such extra Lamb shifts cancel
exactly, order by order in the strength of the driving field, but the
cancellation of course fails if the calculations are not consistent in
the orders. We will not dwell on the latter aspect anymore, but
simply ignore all orders higher than the first one immediately at
the point when the operators have been brought to the desired order.

We illustrate the process of deriving the hierarchy of equations for
operator products with a detailed treatment of the time evolution of a
particular product;
\begin{eqnarray}
\lefteqn{{d\over dt}\,
\left[\psi^\dagger_{g_1}({\bf r}')\psi^\dagger_{g}({\bf r})
\psi_{e}({\bf r})\psi_{g'_1}({\bf r}')\right]}\nonumber\\
&=&
\left[{d\over dt}\,\psi^\dagger_{g_1}({\bf r}')\right]
\psi^\dagger_{g}({\bf r})
\psi_{e}({\bf r})\psi_{g'_1}({\bf r}')+
\psi^\dagger_{g_1}({\bf r}')
\left[{d\over dt}\,\psi^\dagger_{g}({\bf r})\psi_{e}({\bf r})
\right]\psi_{g'_1}({\bf r}')\nonumber\\
&&+ \psi^\dagger_{g_1}({\bf r}')\psi^\dagger_{g}({\bf r})
\psi_{e}({\bf r}){d\over dt}\psi_{g'_1}({\bf r}')\,.
\label{eq:ORI}
\end{eqnarray}

In the way of preparation, let us first note from Eq.~(\ref{eq:FEQ}) and
its hermitian conjugate that the electric displacement may be written
in the form
\begin{equation}
{\bf D}^\pm = {\bf D}^\pm_F + {\bf D}^\pm_S\,,
\end{equation}
where the source term ${\bf D}^\pm_S$ is a normal-ordered
combination of atom fields. We begin our analysis with the third
term on the right-hand side of Eq.~(\ref{eq:ORI}). By virtue of
Eqs.~(\ref{eq:GRF}) and~(\ref{eq:DEL}) we first have
\begin{eqnarray}
\lefteqn{\psi^\dagger_{g_1}({\bf r}')\psi^\dagger_{g}({\bf r})
\psi_{e}({\bf r}){d\over
dt}\psi_{g'_1}({\bf r}')}\nonumber\\
&=&
{i\over\hbar}\,\psi^\dagger_{g_1}({\bf r}')\psi^\dagger_{g}({\bf r})
\psi_{e}({\bf r}){\bf d}_{g'_1e'} \cdot{\bf
E}^-({\bf r}')\psi_{e'}({\bf r}')\nonumber\\ &=&
{i\over\hbar\epsilon_0} \psi^\dagger_{g_1}({\bf
r}')\psi^\dagger_{g}({\bf r})
\psi_{e}({\bf r}){\bf d}_{g'_1e'} \cdot\left[{\bf D}^-({\bf r}')-{\bf
P}^-({\bf r}')\right]\psi_{e'}({\bf r}')
\nonumber\\
&=&{i\over\hbar\epsilon_0}\left[\psi^\dagger_{g_1}({\bf r}')
\psi^\dagger_{g}({\bf r}){\bf d}_{g'_1e'}\cdot{\bf D}^-({\bf
r}')\psi_{e}({\bf r})
\psi_{e'}({\bf r}')-\psi^\dagger_{g_1}({\bf r}')
\psi^\dagger_{g}({\bf r})\psi_{e}({\bf r}){\bf d}_{g'_1e'}\cdot{\bf
P}^-({\bf r}')
\psi_{e'}({\bf r}')
\right]
\nonumber\\
&=&{i\over\hbar\epsilon_0}\left[
\psi^\dagger_{g_1}({\bf r}')
\psi^\dagger_{g}({\bf r}){\bf d}_{g'_1e'}\cdot{\bf
D}_S^-({\bf r}')\psi_{e}({\bf r})
\psi_{e'}({\bf r}')
\right.\nonumber\\
&&+\left.\psi^\dagger_{g_1}({\bf r}')
\psi^\dagger_{g}({\bf r}){\bf d}_{g'_1e'}\cdot{\bf
D}_F^-({\bf r}')\psi_{e}({\bf r})
\psi_{e'}({\bf r}') -\psi^\dagger_{g_1}({\bf r}')
\psi^\dagger_{g}({\bf r})\psi_{e}({\bf r}){\bf d}_{g'_1e'}\cdot{\bf
P}^-({\bf r}')
\psi_{e'}({\bf r}')
\right]\,.
\end{eqnarray}
Since the total displacement ${\bf D}^-$ commutes with all atom
fields, we have first moved it between atom creation and
annihilation operators. The term involving the source field ${\bf
D}^-_S$ is then readily in normal order. Besides, here the source
field term is third order in the perturbation, so it may be omitted.
However, we are not yet done with operator orderings. First, just as
the free-field operator
${\bf D}^+_F$ is profitably moved to the right of atom operators,
the free-field operator ${\bf D}^-_F$ should be transported all the way
to the left. By virtue of Eq.~(\ref{eq:URF}), this leaves behind two
commutator terms. Nevertheless, both the term with ${\bf D}^-_F$
remaining and the commutators are formally third order in the strength
of the driving field, and we ignore them all. Finally,
the term involving ${\bf P}^-$ is not yet in normal order, but it may be
made so easily by using the commutators of the atom fields. The
rearranged term is third order and negligible, but the generated
commutator is first order. We eventually have 
\begin{eqnarray}
\lefteqn{\psi^\dagger_{g_1}({\bf r}')\psi^\dagger_{g}({\bf r})
\psi_{e}({\bf r}){d\over
dt}\psi_{g'_1}({\bf r}')}\nonumber\\ 
&&=-{i\over\hbar\epsilon_0}\,{\bf d}_{g'_1e'}\cdot{\bf
d}_{eg''}\delta({\bf r}-{\bf r}')
\psi^\dagger_{g_1}({\bf r}')\psi^\dagger_{g}({\bf r})\psi_{g''}({\bf
r}')
\psi_{e'}({\bf r}')\,.
\label{eq:UUSI1}
\end{eqnarray}

The same analysis may be carried out with the other two terms in
Eq.~(\ref{eq:ORI}). The first term contributes nothing in the first
order, while the second term gives a homogeneous term proportional
to $i\delta-\gamma$, a driven term proportional to ${\bf D}^+_F$, and
something of a two-atom analog of radiation reaction. The final result
is
\begin{eqnarray}
\lefteqn{{d\over dt}\,
\left[\psi^\dagger_{g_1}({\bf r}')\psi^\dagger_{g}({\bf r})
\psi_{e}({\bf r})\psi_{g'_1}({\bf r}')\right]}\nonumber\\
&=&(i\delta-\gamma)\psi^\dagger_{g_1}({\bf r}')\psi^\dagger_{g}({\bf r})
\psi_{e}({\bf r})\psi_{g'_1}({\bf r}')
+{i\over\hbar\epsilon_0}\left\{
\psi^\dagger_{g_1}({\bf r}')\psi^\dagger_{g}({\bf r})
\psi_{g'}({\bf r})\psi_{g'_1}({\bf r}'){\bf d}_{eg'}\cdot{\bf
D}^+_F({\bf r})
\right.\nonumber\\
&&\left.
+\psi^\dagger_{g_1}({\bf r}')\psi^\dagger_{g}({\bf r})
\psi_{g'}({\bf r})\psi_{e'}({\bf r}'){\bf d}_{eg'} \cdot
\left[\epsilon_0{\bf
S}'({\bf d}_{g'_1e'};{\bf r}-{\bf r}')
-{\bf d}_{g'_1e'}\delta({\bf r}-{\bf r}')\right]\right.\nonumber\\
&&- \psi^\dagger_{g_1}({\bf r}')\psi^\dagger_{g}({\bf r})
{\bf d}_{eg'} \cdot{\bf P}^+({\bf r})
\psi_{g'}({\bf r})\psi_{g_1'}({\bf r}')\nonumber\\
&&+\epsilon_0 \psi^\dagger_{g_1}({\bf r}')\psi^\dagger_{g}({\bf r})
\int d^3r''\,{\bf d}_{eg'} \cdot{\bf S}'({\bf P}^+({\bf r}'');
{\bf r}''-{\bf r})
\left.\psi_{g'}({\bf r})\psi_{g_1'}({\bf r}')
\right\}\,.
\end{eqnarray}

Continuing in this manner, we obtain the equations of motion for an
entire hierarchy of products of atomic operators. As before, we put
the positive-frequency free-field operator to the right, all atom
operators to normal order, and then keep only the terms that are
first order in the perturbation strength. The full result is
\begin{eqnarray}
\lefteqn{{d\over dt}\,
\psi^\dagger_{g_n}({\bf r}_n)\ldots
\psi^\dagger_{g_1}({\bf r}_1)
\psi_{e}({\bf r}_1)\psi_{g'_2}({\bf r}_2)\ldots
\psi_{g'_n}({\bf r}_n)}\nonumber\\
&&=(i\delta-\gamma)\psi^\dagger_{g_n}({\bf r}_n)\ldots
\psi^\dagger_{g_1}({\bf r}_1)
\psi_{e}({\bf r}_1)\psi_{g'_2}({\bf r}_2)\ldots
\psi_{g'_n}({\bf r}_n)\nonumber\\
&&+{i\over\hbar\epsilon_0}\left\{
\psi^\dagger_{g_n}({\bf r}_n)\ldots\psi^\dagger_{g_1}({\bf r}_1)
\psi_{g}({\bf r}_1)\psi_{g'_2}({\bf r}_2)\ldots\psi_{g'_n}({\bf r}_n)
{\bf d}_{eg}\cdot{\bf D}^+_F({\bf r}_1)\right.\nonumber\\
&&-\psi^\dagger_{g_n}({\bf r}_n)\ldots
\psi^\dagger_{g_1}({\bf r}_1){\bf d}_{eg}\cdot{\bf
P}^+({\bf r}_1)\psi_{g}({\bf
r}_1)\psi_{g'_2}({\bf r}_2)\ldots\psi_{g'_n}({\bf r}_n)
\nonumber\\
&&+\epsilon_0\psi^\dagger_{g_n}({\bf r}_n)\ldots
\psi^\dagger_{g_1}({\bf r}_1)\int d^3r'\,{\bf d}_{eg}\cdot{\bf S'}({\bf
P}^+ ({\bf r}');
{\bf r}_1-{\bf
r'})\psi_{g}({\bf r}_1)\psi_{g'_2}({\bf r}_2)\ldots\psi_{g'_n}({\bf
r}_n)
\nonumber\\
&&+\epsilon_0\,\psi^\dagger_{g_n}({\bf r}_n)\ldots
\psi^\dagger_{g_1}({\bf r}_1)\psi_{g}({\bf r}_1)
\sum_{k=2}^n\psi_{g'_2}({\bf r}_2)\ldots\psi_{g'_{k-1}}({\bf r}_{k-1})
\psi_{e'}({\bf r}_k)\psi_{g'_{k+1}}({\bf
r}_{k+1})\ldots\psi_{g'_n}({\bf r}_n)
\nonumber\\
&&\left.\times\,{\bf d}_{eg}\cdot{\bf
W}({\bf d}_{g'_ke'};{\bf r}_1-{\bf r}_k)\right\}\,,
\label{eq:FR2}
\end{eqnarray}
where the notation in the last term implies that $\psi_{g'_k}({\bf
r}_k)$ is missing from the $k$ term on the sum. We have defined
\begin{eqnarray}
{\bf W}(\mbox{\boldmath$\cal D$};{\bf r}) &=& {\bf
S}'(\mbox{\boldmath$\cal D$};{\bf r}) -
{1\over\epsilon_0}\,\mbox{\boldmath$\cal D$}\,\delta({\bf r})
\nonumber\\
&=&{\bf K}(\mbox{\boldmath$\cal D$};{\bf r}) -
{1\over3\epsilon_0}\mbox{\boldmath$\cal D$}\delta({\bf r})\,.
\label{eq:UUSI2}
\end{eqnarray}
This is precisely the classical expression of the electric field (not
displacement) of a dipole $\mbox{\boldmath$\cal D$}$ residing at the
origin, as measured at the point ${\bf r}$. Even the peculiar delta
function divergence of the dipolar field at the origin \cite{JAC75} is
there.

\subsection{A hierarchy for correlation functions}\label{sec:cor}

By taking expectation values of the operator hierarchy
(\ref{eq:FR2}), we obtain a hierarchy of equations of motion for
correlation functions. In order to simplify, in the rest of the paper
we only consider a $J_g=0\rightarrow J_e=1$ transition. Then there are
no Zeeman substates in the ground level, and a single $g$ suffices in
all of the Eqs.~(\ref{eq:FR2}). The three excited Zeeman states are
also handled easily, c.f.\ Appendix \ref{app:VME}.

We define a succession of correlation functions
\begin{eqnarray}
{\bf P}_1(;{\bf r}_1) &=& \langle \psi^\dagger_g({\bf r}_1) {\bf d}_{ge}
\psi_e({\bf r}_1)\rangle \equiv \langle {\bf P}^+({\bf r}_1) \rangle\,,
\nonumber\\
{\bf P}_2({\bf r}_1;{\bf r}_2) &=&
\langle \psi^\dagger_g({\bf r}_1){\bf P}^+({\bf r}_2)
\psi_g({\bf r}_1)\rangle\,,\nonumber\\ {\bf P}_3({\bf r}_1,{\bf
r}_2;{\bf r}_3) &=&
\langle \psi^\dagger_g({\bf r}_1)\psi^\dagger_g({\bf r}_2)
{\bf P}^+({\bf r}_3) \psi_g({\bf r}_2)
\psi_g({\bf r}_1)\rangle\,,\nonumber\\
&...&\label{eq:MPD}\,,
\end{eqnarray}
and similarly
\begin{eqnarray}
\rho_1({\bf r}_1) &=& \langle \psi_g^\dagger({\bf r}_1) \psi_g ({\bf
r}_1)
\rangle\,,
\nonumber\\
\rho_2({\bf r}_1,{\bf r}_2) &=& \langle \psi_g^\dagger({\bf r}_1)
\psi_g^\dagger({\bf r}_2) \psi_g ({\bf r}_2) \psi_g ({\bf r}_1)
\rangle\,,
\nonumber\\
&\ldots&\label{eq:MND}\,.
\end{eqnarray}
${\bf P}_k({\bf r}_1,\ldots,{\bf r}_{k-1};{\bf r}_k)$ is the correlation
function of polarization at ${\bf r}_k$ and atom density at $k-1$
positions
${\bf r}_1,\ldots,{\bf r}_{k-1}$, and $\rho_k$ is a $k$-point
density correlation function. All of these are normally ordered.

We finally reiterate that the driving field is in a coherent state,
so that the factorization (\ref{eq:FCT}) is warranted. In fact, without
further ado, we let ${\bf D}^+_F$ stand for the {\em expectation
value\/} of the coherent free field, or, equally well, for a classical
incident field. It is now a simple matter to derive a hierarchy of
equations for the correlation functions from the operator equations
(\ref{eq:FR2}). We consolidate the results into the form
\begin{mathletters}
\label{eq:HRQ}
\begin{eqnarray}
\dot{\bf P}_1(;{\bf r}_1) &=& (i\delta-\gamma) {\bf P}_1(;{\bf r}_1)
+ i \kappa \rho_1({\bf r}_1) {\bf D}^+_F({\bf r}_1)
+ \int d^3r_2\,{\sf G}({\bf r}_1-{\bf r}_2) {\bf P}_2({\bf r}_1;{\bf
r}_2)\,,
\label{eq:HR1}\\
\dot{\bf P}_2({\bf r}_1;{\bf r}_2)&=&
(i\delta-\gamma) {\bf P}_2({\bf r}_1;{\bf r}_2)
+{\sf G}({\bf r}_2-{\bf r}_1) {\bf P}_2({\bf r}_2;{\bf r}_1)\nonumber\\
&&+i\kappa \rho_2({\bf r}_1,{\bf r}_2) {\bf D}^+_F({\bf r}_2)\nonumber\\
&&+\int d^3r_3\,{\sf G}({\bf r}_2-{\bf r}_3) {\bf
P}_3({\bf r}_1,{\bf r}_2;{\bf r}_3)\,,
\label{eq:HR2}\\
\dot{\bf P}_3({\bf r}_1,{\bf r}_2;{\bf r}_3)&=&
(i\delta-\gamma) {\bf P}_3({\bf r}_1,{\bf r}_2;{\bf r}_3)
+{\sf G}({\bf r}_3-{\bf r}_1) {\bf P}_3({\bf r}_3,{\bf r}_2;{\bf r}_1)
+{\sf G}({\bf r}_3-{\bf r}_2) {\bf P}_3({\bf r}_1,{\bf
r}_3;{\bf r}_2)\nonumber\\ &&+i\kappa \rho_3({\bf r}_1,{\bf r}_2,{\bf
r}_3) {\bf D}^+_F({\bf r}_3)\nonumber\\ &&+\int d^3r_4\,{\sf
G}({\bf r}_3-{\bf r}_4) {\bf P}_4({\bf r}_1,{\bf r}_2,{\bf r}_3;{\bf
r}_4)\,,
\label{eq:HR3}\\
&\ldots&\nonumber\,.
\end{eqnarray}
\end{mathletters}
We have defined the scalar constant
\begin{mathletters}
\begin{equation}
\kappa = {{\cal D}^2\over\hbar\epsilon_0}\,,
\label{eq:NKP}
\end{equation}
and the $3\times3$ tensor
\begin{equation}
{\sf G}_{ij}({\bf r}) = i\kappa\left\{
\left[ {\partial\over\partial r_i}{\partial\over\partial
r_j} - \delta_{ij} \mbox{\boldmath$\nabla$}^2\right] {e^{ikr}\over4\pi
r} -\delta_{ij}\delta({\bf r})
\right\}\,.
\label{eq:GDF}
\end{equation}
\end{mathletters}

The first two equation of the hierarchy (\ref{eq:HRQ}) coincide exactly
with the results of Morice {\it et al.}\cite{MOR95}. These authors
do not proceed any further, but their method, which at this point
had become tantamount to classical electrodynamics, undoubtedly
could have yielded the entire hierarchy.

The terms in (\ref{eq:HRQ}) with $i\delta-\gamma$ obviously come from
the damped free evolution of the polarization in each correlation
function, and the term $\propto{\bf D}^+_F$ corresponds to
excitation of a ground state atom by the driving light to make
polarization. To grasp the dual role of the tensor $\sf G$, let us
consider the equation of motion for
${\bf P}_2({\bf r}_1;{\bf r}_2)$ as an example; correlation function of
polarization at
${\bf r}_2$ and density at ${\bf r}_1$. The integral term obviously
characterizes processes in which a dipole at yet another position
${\bf r}_3$ radiates and thereby promotes a ground state atom at
${\bf r}_2$, so that density becomes dipole density. On the other hand,
the term with
${\sf G}({\bf r}_2-{\bf r}_1)$  describes photon exchange between the
two sites
${\bf r}_1$ and ${\bf r}_2$; an excited atom radiates at ${\bf r}_1$ and
falls to the ground state, while the emitted radiation promotes an atom
at
${\bf r}_2$ to the excited state.

We have implemented several approximations. The most relevant
physical assumption is the perturbative limit with respect to the
strength of the driving light, the most conspicuous technical
assumption is the $J_g=0\rightarrow J_e=1$ transition. However,
Eqs.~(\ref{eq:HRQ}) do not contain any assumptions concerning
spontaneous emission except for our field theory version of the Born and
Markov approximations. Moreover, there are no assumptions, ad hoc or
otherwise, concerning multiple scattering of light or resonant
dipole-dipole interactions; these are included exactly. Hence, so are
collective linewidths and line shifts.

As far as the interactions of atoms with electromagnetic fields are
concerned, we have regarded the atoms as point dipoles. For real
atoms at short distances, when higher multipoles and electron
exchange become relevant, this assumption evidently fails. The
relevance of contact interactions and $\delta$ function contributions
to the dipolar field is questionable, as both operate at {\em zero\/}
distance between the atoms only. Now, real atoms cannot overlap
because of the hard core of the interatomic potential. A reader
troubled by the delta functions may therefore want to consider cutting
off and setting to zero all correlation functions at distances between
the atoms shorter than the typical length scale of a molecular bond.
The effect is that all delta function contributions to the field
propagator ${\sf G}$ of (\ref{eq:GDF}) should be omitted. As the
derivatives of
$e^{ikr}/r$ also produce delta functions, such an omission is
tantamount to replacing the term $-\delta_{ij}\delta({\bf r})$ by
$-{2\over3}\delta_{ij}\delta({\bf r})$. However, in the present paper
we use the propagator exactly as given in Eq.~(\ref{eq:GDF}).

\section{Examples}\label{sec:exa}

In this section we illustrate the correlation function hierarchy with a
few simple examples. At this time we have made little progress toward a
full, exact solution of the hierarchy (\ref{eq:HRQ}) in any
nontrivial situation. Evidently, radically new approaches are needed.
We hope that either ourselves or our readers will in the end be
inspired to come up with a successful solution of, say, the optical
response of a Bose-Einstein condensate (BEC) in the limit of truly
dense sample, $\rho\lambda^3\gg1$.

\subsection{Semi-infinite BEC without collective
coupling}\label{sec:sib}

\subsubsection{Nature of condensate}

An ideal Bose condensate of noninteracting particles is made of a
macroscopic number of particles in the same one-particle quantum
state. Traditionally, the condensate is described by a macroscopic wave
function, whose absolute square gives the particle density and which
also has a phase. More in the vein of quantum optics, one could assume
that the condensate is in a coherent state, albeit with an unknown
phase. In normally ordered operator expressions the field operator
$\psi_g$ then behaves as a c-number. We write
\begin{eqnarray}
\lefteqn{\rho_k({\bf r}_1,\ldots,{\bf r}_k)
= \langle \psi^\dagger_g({\bf r}_1) \ldots\psi_g({\bf r}_1)\rangle}\\
&&\simeq \psi^*_g({\bf r}_1)\ldots\psi^*_g({\bf r}_k)
\psi_g({\bf r}_k)\ldots\psi_g({\bf r}_1)
=
\psi^*_g({\bf
r}_1)\psi_g({\bf r}_1)\ldots\psi^*_g({\bf r}_k)\psi_g({\bf
r}_k)\nonumber\\ &&=
\langle \psi^\dagger_g({\bf r}_1) \psi_g({\bf r}_1)\rangle\ldots \langle
\psi^\dagger_g({\bf r}_k) \psi_g({\bf r}_k)\rangle = \rho_1({\bf
r}_1)\ldots
\rho_1({\bf r}_k)\,.
\end{eqnarray}
In other words, density correlation functions factorize.
For the purposes of the present paper, we take the factorization of
normally ordered density correlation functions as the hallmark of the
condensate even for a weakly interacting condensate. We ignore
noncondensate atoms altogether.

Contrary to the experimental realities, we
ignore the finite dimensions of the condensate. We cannot outright
declare $\rho_1({\bf r})$ as a constant all over space, because
propagation through an infinite medium would cause extinction of all
light before it reaches any  position with finite $|{{\bf r}}|$.
Instead, we assume that a homogeneous condensate with density $\rho$
fills the half-space
$z\ge0$. We assume that the incident and induced radiations as well
as the induced polarizations all propagate in the $z$ direction.
Finally, as we have the $m_g=0$ spherically symmetric ground state
that cannot exhibit any directional preferences, we take all fields
to have the same transverse polarization
$\hat{\bf e}$. In particular, the initial free field is written
\begin{equation}
{\bf D}^+_F({\bf r}) = D_F\,\hat{\bf e}\,e^{ikz}\,.
\label{eq:FFT}
\end{equation}
This has the dispersion relation of light in vacuum, an
oddity in the presence of matter. In fact, in accordance with the
Ewald-Oseen extinction theorem \cite{BOR75}, it will turn out that
the matter responds with a field that exactly cancels the applied
field.

\subsubsection{Optical response without collective coupling}

We solve the response by ignoring the collective line shifts and
dampings, i.e., those $\sf G$ terms in Eqs.~(\ref{eq:HRQ}) that
do not appear inside integrals. It turns out that the simplest
conceivable ansatz, a fully factorized, damped plane wave solution of
the form
\begin{equation}
{\bf P}_n({\bf r}_1,\ldots;{\bf r}_n) = 
\left\{
\begin{array}{ll}P\,\rho^{n-1}\,\hat{\bf e}\,e^{ik'z_n},&z_1\ge0,
\ldots z_n\ge0;\\ 0, &{\rm otherwise}.
\end{array}
\right.
\label{eq:NCS}
\end{equation}
succeeds for suitable choices of $P$ and $k'$, with $\Im(k') >
0$.

To see this, we need the integral
\begin{equation}
\int_{z_2\ge0}d^3r_2\,\hat{\bf e}^* \cdot{\sf
G}({\bf r}_1-{\bf r}_2) \cdot\hat{\bf e}\, e^{ik'z_2}
= i\kappa\left[{k^2\over k'^2-k^2}\,e^{ik'z_1} +
{{k'}^2\over2k(k-k')}\,e^{ikz_1}\right]\,,
\label{eq:PRI}
\end{equation}
where the vector $\hat{\bf e}$ takes care of the polarizations. The
integral is valid for $z_1>0$. For $z_1<0$ the integral yields
a reflected wave $\propto e^{-ikz}$ instead, but we do not consider this
case any further. With the ansatz (\ref{eq:NCS}), the integrals on the
right hand sides of (\ref{eq:HRQ}) produce sums of two exponentials, one
with the wave number $k$ appropriate for light in vacuum and the other
with the wave number
$k'$ for light in the medium. In steady state of Eqs.~(\ref{eq:HRQ}) the
vacuum component $\propto e^{ikz}$ must cancel the corresponding
free-field terms, and the remaining $e^{ik'z}$ term must pair up with
the polarization correlation functions ${\bf P}_n$. All of the
Eqs.~(\ref{eq:HRQ}) then reduce to the following two conditions:
\begin{mathletters}\label{eq:SSC}
\begin{eqnarray}
\left(i\delta-\gamma + {i\rho\kappa k^2\over k'^2-k^2} \right)P &=& 0
\,,\label{eq:SS1}\\
D_F + {{k'}^2\over 2k(k-k')}\,P &=& 0\,.
\label{eq:SS2}
\end{eqnarray}
\end{mathletters}
The first one gives the wave number $k'$, and the second one may then
be read as a condition for the polarization amplitude $P$.

One expects that the total displacement from (\ref{eq:FEQ}) should be
of the form $D\,\hat{\bf e}\,e^{ik'z}$ as well. With the choice
(\ref{eq:SS2}), the vacuum type contributions $e^{ikz}$ indeed cancel.
We have the condition for the polarization amplitude and the amplitude
of electric displacement,
\begin{equation}
D = {k'^2\over k'^2-k^2}\,P\,.
\label{eq:DVP}
\end{equation}
This implies that an electric field of the form
$E\,\hat{\bf e}\,e^{ik'z}$ also propagates in the medium, with the
amplitude given by
\begin{equation}
E = {D-P\over\epsilon_0} = {1\over\epsilon_0}\,{k^2\over k'^2-k^2}\,P\,.
\label{eq:EFP}
\end{equation}
As customary, we define refractive index $n$ in such a way that
$k' = nk$, and susceptibility $\chi$ such that $P = \epsilon_0\chi E$.
Equations (\ref{eq:SSC})--(\ref{eq:EFP}) immediately give the explicit
expressions
\begin{equation}
n^2 - 1 = \chi = -{\rho\kappa\over\delta+i\gamma}
\label{eq:CHI}
\end{equation}
for these quantities.

The remarkable feature of the result (\ref{eq:CHI}) is that it is so
unremarkable: susceptibility is obtained as atomic
polarizability times atom density. This is precisely the conventional
column density approach that the experimenters routinely use to
analyze their BEC results. In other words, we have proven that, for a
plausible model of the BEC and within a precisely formulated
approximation that ignores collective linewidths and shifts, the
column density approach is {\em exact}. We regard this as an
important result, in that it displays precisely and explicitly the
underlying assumptions of the column density arguments. The flip
side is that at high density simply ignoring the collective effects is
another uncontrolled approximation.

\subsection{Density expansion to second order}\label{sec:det}

In the present section we review the density expansion of Morice {\em
et al}. \cite{MOR95} from the point of view of our development. In
effect, they truncate the hierarchy for correlation functions by writing
\begin{equation}
{\bf P}_3({\bf r}_1,{\bf r}_2;{\bf r}_3) \simeq
{\rho_2({\bf r}_1,{\bf r}_2)\over\rho({\bf r}_2)}
\,{\bf P}_2({\bf r}_2;{\bf r}_3)\,.
\label{eq:SPF}
\end{equation}
If (and probably in some fairly strong sense only if) one inserts this
particular factorization into (\ref{eq:HR2}), one may eliminate
at the same time both the free-field term and the integral term from
the steady-state versions of Eqs.~(\ref{eq:HR1})
and~(\ref{eq:HR2}). This gives an algebraic relation between ${\bf
P}_1$ and ${\bf P}_2$,
\begin{equation}
0 = {\bf P}_2({\bf r}_1;{\bf r}_2) - {{\sf
G}({\bf r}_2-{\bf r}_1)\over(i\delta-\gamma)}\, {\bf P}_2({\bf
r}_2;{\bf r}_1) + {\rho_2({\bf r}_1,{\bf r}_2)\over\rho_1({\bf
r}_2)}\,{\bf P}_1(;{\bf r}_2)\,.
\label{eq:TM4}
\end{equation}

Now assume a constant density of atoms $\rho$ in the half-space
$z\ge0$, and write the physically justifiable ansatz
\begin{equation}
\rho_2({\bf r}_1,{\bf r}_2) = \rho^2\,[1+\varphi({\bf r}_1-{\bf
r}_2)]\,.
\label{eq:DCF}
\end{equation}
The relation (\ref{eq:TM4}) and its counterpart with ${\bf r}_1$
and ${\bf r}_2$ interchanged may then be solved to give
\begin{eqnarray}
{\bf P}_2({\bf r}_1;{\bf r}_2) &=& {\rho[1+\varphi({\bf r}_1-{\bf r}_2)]
\over
1-[{\sf G}({\bf r}_1-{\bf r}_2)/(i\delta-\gamma)]^2}
\left[
{\bf P}_1(;{\bf r}_2) - {{\sf G}({\bf r}_1-{\bf
r}_2)\over(i\delta-\gamma)}\, {\bf P}_1(;{\bf r}_1)
\right]\nonumber\\
&=& \rho{\bf P}_1(;{\bf r}_2) +{\rho\over1-g^2}\,
[(\varphi\!+\!g^2){\bf P}_1(;{\bf r}_2) - g{\bf P}_1(;{\bf r}_1)]\,,
\label{eq:PP2}
\end{eqnarray}
where we have introduced an obvious temporary notation. The reason for
the split in the second form of (\ref{eq:PP2}) is the following. Suppose
we attempt an ansatz of the form (\ref{eq:FFT}), together with
${\bf P}_1(;{\bf r}) = P\,\hat{\bf e}\,e^{ik'z}$. Then, when
(\ref{eq:PP2}) is inserted into the integral in right-hand side of
(\ref{eq:HR1}), in accordance with (\ref{eq:PRI}) the term
$\rho {\bf P}_1$ will produce a term with the
spatial dependence of the free field $e^{ikz}$. All other
contributions to the integral will behave as $e^{ik'z}$, at least for
$z\rightarrow\infty$. Now, the requirement that the free-field
contributions cancel (Ewald-Oseen extinction theorem \cite{BOR75}
again) gives one relation between $D_F$, $P$ and $k'$, and
Eq.~(\ref{eq:FEX}) furnishes another. One then obtains an equation
out of which one may solve $k'$, and therefore ultimately the
entire response of (a thick slab) of the gas. This equation reads
\begin{mathletters}
\begin{equation}
{{k'}^2\over k^2} = 1 -
{\kappa\rho\over\delta+i\gamma}\,{1\over1+C}\,,
\end{equation}
with
\begin{eqnarray}
C &=& {\rho\over i\delta-\gamma}\int d^3r\,\varphi({\bf r})e^{-ik'z}
\,\hat{\bf e}^* \cdot{\sf G}({\bf r}) \cdot\hat{\bf e}\nonumber\\
&&+\rho\int d^3r\,[1+\varphi({\bf r})]\,\hat{\bf e}^* \cdot\left[
{e^{-ik'z}[{\sf G}({\bf r})/(i\delta-\gamma)]^3 - [{\sf
G}({\bf r})/(i\delta-\gamma)]^2
\over
1 - [{\sf G}({\bf r})/(i\delta-\gamma)]^2}
\right] \cdot\hat{\bf e}\,.
\end{eqnarray}
\end{mathletters}

Ingenuous as the analysis of Ref.~\cite{MOR95} is, it elicits two
questions. First, the factorization (\ref{eq:SPF}) is the way it is to
facilitate the mathematics, not because (\ref{eq:SPF}) would be a
particularly apt physics assumption. There is no guarantee that the
correlations between the ground state atoms are treated adequately.
Second, while Morice {\it et al}.\ argue that the result is the correct
expansion in the parameter $\rho\lambda^3$ up to the order
$(\rho\lambda^3)^2$ and, in fact, includes all multiple scattering
events between any {\em pair} of atoms, the mathematical structure of
the hierarchy (\ref{eq:HRQ}) does not directly bear this out. The
difficulty is that each integral of the form  
$\int d^3r_k\,{\sf G}{\bf P}_{k}$ must produce a component that
cancels the associated free-field term. In other words, the integral
does not automatically signal an increasing power in $\rho\lambda^3$. 
Within the present approach, a rigorous mathematical
counting of the powers of $\rho\lambda^3$ seems elusive.

\subsection{Resonant dipole-dipole interaction of two
atoms}\label{sec:rdd}

Let us take two atoms with small non-overlapping c.m.\ wave packets
$\phi_\pm({\bf r})$ centered around ${\bf r}_\pm$. Inasmuch as the
sizes of the wave packets are much smaller than the wavelength of the
exciting light, for the purposes of the analysis of optical response 
we may write the atom density and the density correlation function as
\begin{eqnarray}
\rho_1({\bf r}_1) &=& |\phi_+({\bf r}_1)|^2 + |\phi_-({\bf r}_1)|^2
\simeq \delta({\bf r}_1-{\bf r}_+) + \delta({\bf
r}_1-{\bf r}_-)\,,\nonumber\\
\rho_2({\bf r}_1,{\bf r}_2) &=& |\phi_+({\bf r}_1)|^2 |\phi_-({\bf
r}_2)|^2 
	+ |\phi_+({\bf r}_2)|^2 |\phi_-({\bf r}_1)|^2\nonumber\\
&\simeq&\delta({\bf r}_1-{\bf r}_+)\delta({\bf r}_2-{\bf r}_-)
+\delta({\bf r}_2-{\bf r}_+)\delta({\bf r}_1-{\bf r}_-)\,.
\end{eqnarray}
To simplify the results further we assume that the geometry of the
situation is such that for the driving light at the positions of the
atoms we have ${\bf D}^+_F({\bf r}_+)= {\bf D}^+_F({\bf r}_-)\equiv 
{\bf D}^+_F$. In particular, this is justified if the atoms are well
within a wavelength of one another. In such a case the two-atom system
may be discussed in terms of Dicke states, some superradiant and some
subradiant. The field configuration we have chosen does not excite
the subradiant states at all, so these will not come up in our
analysis.

Since we have two atoms present, all correlation functions referring to
more than two atoms vanish: $\rho_n = 0$ and ${\bf P}_n = 0$
for $n = 3,4,\ldots$. The steady state of the hierarchy (\ref{eq:HRQ})
is then found trivially:
\begin{equation}
{\bf P}_1(;{\bf r}_1) = {\bf p}[\delta({\bf r}_1-{\bf r}_+) +
\delta({\bf r}_1-{\bf r}_-)]\,,
\end{equation}
with
\begin{equation}
{\bf p} = - {i\kappa\over(i\delta-\gamma)+{\sf G}}\,{\bf D}^+_F\,.
\label{eq:RES}
\end{equation}

Unraveling Eq.~(\ref{eq:RES}) (with the $3\times3$ tensor $\sf G$ in
the denominator) gives a complicated expression that depends on the
polarization of the driving light ${\bf D}^+_F$, direction between the
atoms, and the distance between the atoms $r$. The main features,
however, are intuitive obvious. For $r\ll\lambda$ the optical response
shows a split resonance at
\begin{equation}
\delta = {\kappa\over4\pi r^3},\qquad
\delta = -{\kappa\over2\pi r^3}\,.
\label{eq:SPL}
\end{equation}
The configurations that produce only one or the other of these
resonance are such that the polarization of the driving field (and
hence, of the induced dipoles) is perpendicular or parallel to the
vector joining the dipoles, respectively. These correspond the doubly
degenerate $\pi_u$ and the nondegenerate $\sigma_u$ configurations of
the molecule consisting of the two atoms. The cooperative width of the
resonances is $2\gamma$  instead of the one-atom linewidth
$\gamma$. Incidentally, the subradiant states that are invisible
within our approximations correspond to the molecular configurations 
$\pi_g$ and $\sigma_g$, which appropriately do not have dipole
coupling to the molecular ground state.

Because our theory is linear in the external field, it should not come
as a surprise that exactly the same dipole-dipole response ensues
for two classical, isotropic, charged harmonic oscillators. In such a
calculation one may want to put in the one-oscillator damping
in $\gamma$ by hand in order to dispense with an explicit treatment of
the divergent radiation reaction. Nonetheless, the collective linewidth
and dipole-dipole interactions are easily derived from the classical
analysis of the radiation that the oscillators exchange among
themselves. On the other hand, the molecular analogy also hints at some
aspects of physics that are missing from our formulation. First, once
more, at close enough distances two atoms may no longer be
regarded as point dipoles. Second, since we have ignored all c.m.\
evolution, the possibility of quantized vibrational states of the
molecules and the ensuing optical resonances have fallen by the
wayside.

As the connection of our hierarchy to dipole-dipole interactions
has now come up explicitly, we should point out that there are
numerous discussion of two- and $n$-atom response in the literature
in which (in effect) the dipole-dipole interaction is derived (in
effect) by eliminating vacuum electromagnetic fields with the aid of
Born and Markov approximations; see, e.g., \cite{LEH70}. As it comes
to dense and/or degenerate gases, it seems that dipole-dipole
interactions are going to occupy an increasingly prominent position
in theory and eventually perhaps in experiments as well. ``Nonlinear
atom optics'' \cite{LEN93,ZHA94,LEN94,ZHA95}, in which the
self-interaction of the atom wave derives from dipole-dipole
interactions, is a prominent example. In fact, the analysis of Zhang
and Walls
\cite{ZHA94} in terms of Heisenberg picture field operators is in
spirit quite close to our treatment, and it seems plausible that our
correlation function hierarchy could also be derived from the
Schr\"{o}dinger picture master equation as presented by Lenz {\em et
al\/} \cite{LEN94}.

\subsection{Prospects for exact solution}\label{sec:pfe}

Dipole radiation presents mathematical difficulties both at short and
long distances. One has the $1/r^3$ and indeed a delta function
divergence at short distances, which means that the results are
sensitive to short range correlations between the atoms. Also, the
dipole interaction falls off as $1/r$ at large distances. Integrals
involving the dipole interaction are not absolutely convergent on the
fall-off of the interaction alone, and local approximations of the type
\[
\int d^3r'\,{\sf G}({\bf r}-{\bf r}')f({\bf r}')
\simeq
f({\bf r}) \int d^3r'\,{\sf G}({\bf r}-{\bf r}')
\]
cannot be made. There may be a global coupling between the
electromagnetic fields and polarization reaching across the entire
sample. At the face of such mathematical hazards, any uncontrolled
approximation should be viewed with suspicion. An essentially
exact, most likely numerical, solution of the hierarchy (\ref{eq:HRQ})
appears highly desirable.

The idea of attempting a direct solution of the hierarchy
of integral equations (\ref{eq:HRQ}) in any conceivable future is
clearly stillborn. Instead, on several occasions we have seen hints for
an alternative. We have noted in our treatment of two atoms that in the
limit of low light intensity the atoms act like classical charged
harmonic oscillators. Moreover, even though the derivation was in terms
of commutator properties of various fields, both the shift and the
damping of the atoms associated with the dipole-dipole interaction
could be viewed simply as manifestations of the classical radiation
transmitted from one atom to the other. It is conceivable that a proper
(stochastic) spatial distribution of classical radiators could share
the correlation function hierarchy (\ref{eq:HRQ}). The hierarchy might
thus be solved by simulating a system of classical atoms and classical
electromagnetic fields numerically. However, so far we have no
mathematically rigorous prescription for such a simulation, let alone
a practical implementation.

\section{Concluding remarks}\label{sec:con}

We have presented a fully quantum mechanical, careful analysis
of the response of a gas, possibly degenerate, to electromagnetic
fields. The main technical ingredients are the field theory version
of Born and Markov approximations, and procedures to move atom
fields and electromagnetic fields to a certain (basically normal)
order.  The outcome is a hierarchy of equations of motion for atomic
correlation functions, in this paper specifically developed for the
limit of low light intensity.

Under our assumptions, notably low intensity, the hierarchy 
is in a sense obvious. In retrospect, it could have been outright
guessed on the basis of classical electrodynamics. We note, though,
that our methods would work in many generalizations that go beyond
classical physics. Given that even the simplest low-intensity limit
has not yet been solved satisfactorily for near resonance response 
of a dense gas, we do not address more complicated cases at
any length. Nonetheless, a few possible generalizations should be
mentioned.

The price for an arbitrary intensity would be a more complicated
hierarchy containing atomic correlation functions with more than
one excited state field $\psi_e$. In the case of an arbitrary
intensity, a two-level atom no longer behaves identically to a
charged harmonic oscillator, scattered light may be nonclassical,
and statistics of the atoms may play a nontrivial role. The
hierarchy for an arbitrary intensity most likely will not admit a
classical simulation. Solutions for a finite number of atoms and
density expansions might be extracted, but the full hierarchy would
present a truly daunting problem.  

While we have treated the c.m.\ degrees of freedom quantum
mechanically and at all times properly retained the quantum
statistics of the atoms, we have for the most part ignored the c.m.\
Hamiltonian. We have effectively consigned the atoms to immobility.
This is by no means necessary. We could add kinetic energy of the
atoms, a confining potential, and even molecular potential curves to
the theory. Of course, this again entails complications: the entire
physics of molecules made of a single atomic species, photon
recoil, cold collisions, etc., become special cases of our approach.

The ultimate objective of our formulation is to find the
expectation value of the polarization, $\langle {\bf P}^+\rangle$.
Out of $\langle {\bf P}^+\rangle$ one may deduce the expectation
value of the scattered field, and hence, the expectation value of
the total electric field $\langle {\bf E}^+\rangle$. The flaw here
is that a typical detector of light does not measure
$\langle {\bf E}^+\rangle$, but rather expectation values of quadratic
quantities such as $\langle {\bf E}^-{\bf E}^+\rangle$. One may
approximate, say, the measured intensity as
\begin{equation}
I({\bf r} t) = \langle {\bf E}^-({\bf r} t) {\bf E}^+({\bf r} t)\rangle
\simeq
\langle {\bf E}^-({\bf r} t)\rangle \langle {\bf E}^+({\bf r}
t)\rangle\,,
\end{equation}
but it is known in quantum optics that this type
of an approximation broadly speaking misses the intensity of {\em
inelastically} scattered light. Our hierarchy is tantamount to a
collection of classical linear harmonic oscillators interacting
with light, a system in which one expects elastic scattering only.
However, when one goes beyond the low-intensity limit,
even a single two-level atom scatters inelastically; and if the
c.m.\ hamiltonian is fully included, photon recoil gives
additional inelastic scattering. To include inelastic scattering
properly, one needs to develop a hierarchy starting from
$\langle {\bf P}^-({\bf r} t) {\bf P}^+({\bf r}' t) 
\rangle$, and proceed consistently at least in the {\em
second} order in the strength of the driving electric field.

Finally, to calculate the spectrum of scattered radiation, one
employs the two-time correlation
function of the electric field
$\langle {\bf E}^-({\bf r} t) {\bf E}^+({\bf r} t')\rangle$,
which is obtainable from the two-time polarization correlation
function $\langle {\bf P}^-({\bf r} t) {\bf P}^+({\bf r}' t') \rangle$.
To compute the latter, one needs not only the field theory version of
Born and Markov approximations, but also further considerations
that essentially amount to the regression theorem \cite{MEY91}.
 
The modifications of the spectrum of the scattered light in the
presence of the condensate discussed in Ref.~\cite{JAV95} 
are due to inelastic scattering associated with photon recoil.
Ironically, to analyze even this seemingly simple case within
our present framework, we would have to generalize so that the c.m.\
motion is included, and also develop the quantum regression theorem.
Our linear hierarchy is no panacea; major generalization
are needed in many relevant problems. Nonetheless, we hope that the
eventual solutions of the hierarchy, by means of classical
simulations or otherwise, will shed new light on the near-resonance
optical response of dense atomic samples.

\acknowledgements
We thank Carl Williams for advice regarding molecular physics. This
work is supported in part by the National Science Foundation.

\appendix
\section{Mathematical details}\label{app:MDT}
\subsection{Dipole radiation with wave number cutoff}\label{app:CMT}

We first consider the following integral involving the propagator that
governs dipole radiation from matter fields:
\begin{equation}
I(\rho) = \int_0^\infty dt\,\int_{r<\rho} d^3r\, {\bf
S}(\mbox{\boldmath$\cal D$};{\bf r},t)\,.
\label{eqIT1}
\end{equation}
Here the ${\bf r}$ integral runs over a sphere of radius $\rho$. Given
the finite-width delta function (\ref{eq:DWT}), we have 
\begin{eqnarray}
I(\rho) &=&
-{1\over4\pi\epsilon_0}\int_{r<\rho}d^3r\,
\mbox{\boldmath$\nabla$}\times(\mbox{\boldmath$\cal
D$}\times\mbox{\boldmath$\nabla$}) {\hbox{\rm erf}(r/\alpha)\over
r}\nonumber\\ &=& -{1\over4\pi\epsilon_0} \int_{S} dS\,
\hat{\bf n}\times\left[
\mbox{\boldmath$\cal D$}\times\mbox{\boldmath$\nabla$}\,{\hbox{\rm
erf}(r/\alpha)\over r}\right]\,,
\end{eqnarray}
where the integral now runs over the surface $S$ of the sphere.

Two obvious limiting cases emerge depending on the radius of the
sphere. For $\rho\gg\alpha$ the error function inside the integral
may be regarded as a constant equal to one, and the integral gives
\begin{equation}
I(\rho) = {2\over3\epsilon_0}\,\mbox{\boldmath$\cal D$}
,\qquad\rho\gg\alpha\,.
\label{eq:ITB}
\end{equation}
In the contrary case $\rho\ll\alpha$ we may expand the error function
as a power series in $r$. This immediately yields
\begin{equation}
\lim_{\rho\rightarrow0} I(\rho) = 0\,.
\label{eq:ITC}
\end{equation}
We have, in effect, shown that there is something akin to a
delta function, albeit with a finite width $\sim\alpha$, in the
immediate proximity of the origin ${\bf r} = 0$ in the expression $\int
dt\,{\bf S}$.

Next consider an integral of the form
\begin{equation}
I = \int_0^\infty dt\,\int d^3r'\, {\bf S}(\mbox{\boldmath$\cal D$},
{\bf r}-{\bf r'},t-t')
\phi({\bf r'} t') e^{-i\Omega (t-t')}\,,
\label{eqIT7}
\end{equation}
where the characteristic spatial and temporal scales
of the function $\phi$ satisfy
\begin{equation}
\Delta r\gg \alpha,\quad\Delta t \gg \alpha/c,\qquad \Delta t \gg
\Omega^{-1}\,.
\end{equation}
Moreover, we assume that the cutoff of electromagnetic frequencies is
much higher than $\Omega$, $c/\alpha\gg\Omega$. With these conditions,
it is possible to choose a length $\rho$ such that
simultaneously $\alpha\ll\rho\ll \Delta r$ and $\rho\ll
c\Omega^{-1} =
\lambda/2\pi$. We divide the spatial integral into two regions: a
sphere of radius $\rho$, and the complement of the sphere. In the outer
region the function $\delta_\alpha$ acts effectively as a delta
function, so we have
\begin{eqnarray}
I_{r\ge\rho} &=& {1\over4\pi\epsilon_0}\int_{r\ge\rho} d^3r'\,
\mbox{\boldmath$\nabla$}\times(\mbox{\boldmath$\cal
D$}\times\mbox{\boldmath$\nabla$})
\left[{\phi({\bf r'},t-|{\bf r}-{\bf r'}|/c)e^{ik|{\bf r}-{\bf r'}|}
\over|{\bf r}-{\bf r'}|}\right]\nonumber\\
&\simeq&\int_{r\ge\rho} d^3r'\,
{\bf K}(\mbox{\boldmath$\cal D$};{\bf r}-{\bf r'}) 
\phi({\bf r'},t-|{\bf r}-{\bf r'}|/c)\,,
\end{eqnarray}
where {\bf K} is the standard dipole radiation formula
(\ref{eq:DOL}). On the other hand, in analogy with the previous
expression (\ref{eq:ITB}), the integral over the inner region gives
\begin{equation}
I_{r<\rho} = {2\over3\epsilon_0}\,\mbox{\boldmath$\cal D$}\,\phi({\bf r}
t)\,.
\end{equation}

Summarizing, when the result is used in conjunction with a smooth
function of ${\bf r}$ and $t$ that also varies slowly in comparison with
$e^{-i\Omega t}$, we may write
\begin{eqnarray}
\int_0^{\infty}dt\,{\bf S}(\mbox{\boldmath$\cal D$};{\bf
r},t)e^{-i\Omega t} &=& {1\over4\pi\epsilon_0}(\mbox{\boldmath$\cal
D$}\times\mbox{\boldmath$\nabla$})\times\mbox{\boldmath$\nabla$})
{e^{ikr}\over r}
\nonumber\\
&=& {1\over4\pi\epsilon_0}
\left[{\bf K}(\mbox{\boldmath$\cal D$},{\bf r}) +
{8\pi\over3}\,\mbox{\boldmath$\cal D$}\,\delta({\bf r})\right]\,.
\end{eqnarray}
Besides, this prescription comes with explicit directions how to
handle integrals involving the dipole radiation. Because of the $1/r^3$
divergence of the dipole radiation $\bf K$, such integrals
are generally not absolutely convergent, and their values
depend on how they are done. It is clear from our development that the
proper way to carry out such integrals is to remove a sphere of a
finite radius $\rho$ around the divergence, calculate the integral,
and then let $\rho\rightarrow0$. In practice, this is the same as
doing the integral in spherical coordinates with the origin at the
divergence, and integrating over the angles first. This is the
prescription adopted in the present paper.

One may wonder where precisely our rule for handling the $1/r^3$
singularity came from, and whether there are plausible alternatives.
Mathematically, our rule originates from the assumption that the
photon  modes were truncated in a manner that preserves the isotropy of
photon phase space. We surmise, albeit without proof, that our
prescription, indeed, is essentially unique if there is to be no
intrinsically favored directions for photons.  

\subsection{Vectors and matrix elements}\label{app:VME} Alongside
with the Cartesian unit vectors, we introduce the conventional
circular unit vectors as
\begin{equation}
\hat{\bf e}_+=-\frac{1}{\sqrt{2}}(\hat{\bf e}_1+\hat{\bf e}_2),\quad
\hat{\bf e}_0 = \hat{\bf e}_3,
\quad\hat{\bf e}_-=\frac{1}{\sqrt{2}}(\hat{\bf e}_1-i\hat{\bf e}_2)\,,
\label{eqa:UNV}
\end{equation} These are orthonormal, in that
\begin{equation}
\hat{\bf e}^*_\sigma\cdot\hat{\bf e}_{\sigma'} = \delta_{\sigma\sigma'},
\quad \sum_\sigma \hat{\bf e}_\sigma\,\hat{\bf e}^*_\sigma = 1\,.
\label{eqa:ORN}
\end{equation} We also define a shorthand for the Clebsch-Gordan
coefficients,
\begin{equation}
\langle J_eM;1J_g|1\sigma J_gm\rangle
\equiv \langle M| m \sigma \rangle
\equiv \langle m \sigma| M \rangle\,.
\label{eqa:CGC}
\end{equation} The dipole operator is defined as
\begin{equation}
{\bf d} = {\cal D} \sum_{mM\sigma} 
|J_eM\rangle \langle M | m\sigma\rangle \langle J_gm | \,\hat{\bf
e}^*_\sigma +
\hbox{H.c.}\,,
\label{eqa:DOP}
\end{equation}
where ${\cal D}$ is the reduced dipole moment matrix element that would
pertain to a transition with unit Clebsch-Gordan coefficient.
In this paper we have chosen ${\cal D}$ to be real. The reason for
the complex conjugate in Eq.~(\ref{eqa:DOP}) is that we want a 
light field with the polarization $\hat{\bf e}_+$ to drive
transitions with $M-m=1$. The dipole matrix elements are explicitly
\begin{equation}
{\bf d}_{Mm} = {\cal D} \sum_\sigma \langle M | m\sigma\rangle\,\hat{\bf
e}^*_\sigma,
\quad {\bf d}_{mM} = {\bf d}_{Mm}^*\,.
\label{eqa:MEL}
\end{equation}
In particular, it may be verified from the orthonormality
of the Clebsch-Gordan coefficient that these matrix elements satisfy
\begin{equation}
\sum_m {\bf d}_{Mm}\cdot{\bf d}_{mM'} = {\cal D}^2 \delta_{MM'}\,.
\label{eqa:DSM}
\end{equation}
In the special case $J_g=0\rightarrow J_e=1$ we have
\begin{equation}
\langle -1 | 0-1\rangle =
\langle 0 |00\rangle =
\langle +1 | 0+1\rangle =1\,.
\label{eq:01G}
\end{equation}

\end{document}